\documentclass[conference]{IEEEtran}
\IEEEoverridecommandlockouts

\usepackage{cite}
\usepackage{amsmath,amssymb,amsfonts}
\usepackage{algorithmic}
\usepackage{graphicx}
\usepackage{textcomp}
\usepackage{xcolor}
\usepackage{xspace}
\usepackage{tabularray}
\usepackage{booktabs}
\usepackage{subfigure}
\usepackage{array}
\usepackage{threeparttable}
\usepackage{makecell}
\usepackage{bbding}
\usepackage[hidelinks]{hyperref}

\def\BibTeX{{\rm B\kern-.05em{\sc i\kern-.025em b}\kern-.08em
    T\kern-.1667em\lower.7ex\hbox{E}\kern-.125emX}}
\begin{document}

\title{CDN Tsunami: Exploiting HTTP/3-HTTP/1.1 Conversion for DoS Attacks}

\author{
\IEEEauthorblockN{Ziyu Lin\IEEEauthorrefmark{1}\IEEEauthorrefmark{2}, Tianlong Su\IEEEauthorrefmark{2}, Yingjie Lin\IEEEauthorrefmark{2}, Prosanta Gope\IEEEauthorrefmark{3}, Yinzhi Cao\IEEEauthorrefmark{4}, Ximeng Liu\IEEEauthorrefmark{2}, Biplab Sikdar\IEEEauthorrefmark{1}}
\IEEEauthorblockA{
\IEEEauthorrefmark{1}National University of Singapore,\quad
\IEEEauthorrefmark{2}Fuzhou University,\quad
\IEEEauthorrefmark{3}University of Sheffield,\quad
\IEEEauthorrefmark{4}Johns Hopkins University}
}

\maketitle

\pagestyle{empty}
 
\begin{abstract}

Content Delivery Networks (CDNs) provide high availability, accelerate content delivery for their host websites, but are also vulnerable to different types of Denial-of-Service (DoS) attacks. Prior works have studied a variety of DoS attacks with HTTP/1.1 or HTTP/2 connections, but most of them are being fixed, making CDNs robust against such attacks. One unexplored research area is how the recent introduction of HTTP/3 at CDNs affects the DoS attack landscape, especially when there are heterogeneous deployments of HTTP/3 and HTTP/1.1 between CDNs and host websites. 

In this paper, we design the \emph{first} study of DoS attacks against HTTP/3 protocols deployed at CDNs. Our key insight is that when the CDN adopts HTTP/3 but the host websites use HTTP/1.1, an adversary can utilize the disparity to amplify a small amount of traffic to the CDN using HTTP/3 to a large amount from the CDN to the host website using HTTP/1.1. More specifically, we design two attack variations---HTTP/3 Bandwidth Amplification (HBA) and HTTP/3 Connection Amplification (HCA)---targeting the bandwidth and the number of connections, respectively. 
 Furthermore, we conduct a large-scale measurement upon the Tranco Top 1M domain list to quantify the real-world impact of these attacks, identifying 42,330 subdomains that are potentially vulnerable to our attacks.  Finally, we responsibly disclose the details of our attacks to the affected CDN vendors: So far, two vendors have already acknowledged their vulnerabilities with bounties and have deployed our mitigations.
\end{abstract}

\section{Introduction}\label{sec: introduction}
Content Delivery Networks (CDNs) are indispensable for the contemporary Internet ecosystem. They significantly reduce user access latency and enhance the user experience by caching content on globally distributed edge servers. A core advantage of CDNs lies in their load-balancing capabilities, which effectively manage massive concurrent requests and ensure global access to content. For example, over 55\% of the Top 10K websites and more than 40\% of the Top 1M websites are deployed behind CDNs~\cite{CDN_Usage_Statistics}. While CDNs provide convenience for real-world websites, they have also become a popular target for attacks, especially Denial-of-Service (DoS) attacks. 

In the past, researchers have proposed many DoS attacks~\cite{2009_decopule_attack,cdn_loop,RangeAmp,CDN-Convex,lin2024unveiling,cdn_cannon} against CDN services. For example, Guo et al.~\cite{cdn_judo} designed an amplification attack that abuses the conversion between HTTP/2 and HTTP/1.1 requests. Li et al.~\cite{RangeAmp} utilize HTTP Range Requests for amplifying the request to the host website with a small request. Triukose et al.~\cite{2009_decopule_attack} proposed an attack that exhausts the host website's bandwidth by quickly disconnecting the client-CDN connection.
However, most existing attacks are widely known and have thus been protected against by existing CDN providers~\cite{akamai_MYR}. 
Furthermore, although prior works have studied various DoS attacks related to different HTTP versions from HTTP/1.1 to HTTP/2, one unexplored research area is the combination of HTTP/3 and CDN services. The HTTP/3 protocol was officially standardized through RFC 9114~\cite{rfc9114} and RFC 9204~\cite{rfc9204} and is currently supported by 35.2\% of websites~\cite{http/3_usage} and most CDN services. 

In this paper, we propose the first study of DoS attacks against HTTP/3 services adopted by CDN providers. The key insight of the attacks is the mismatch between the efficient transmission in terms of bandwidth and number of connections of HTTP/3 and the low efficiency of HTTP/1.1. Therefore, an attacker can establish efficient HTTP/3 connections to the CDN server, which will be expanded to inefficient requests, such as a large number of requests or requests with large payloads, to the host website, leading to a DoS attack.  Specifically, we describe two types of HTTP/3 related DoS attacks: 
\begin{itemize}
\item {HTTP/3 Bandwidth Amplification (HBA) Attack.}
The primary objective of this attack is to exploit the CDN's HTTP/3-to-HTTP/1.1 conversion, where small QPACK index values, a new feature introduced in HTTP/3, are decoded into large raw HTTP headers, thereby launching bandwidth amplification attacks against host websites.
\item {{HTTP/3 Connection Amplification (HCA) Attack}.}  The primary objective of this attack is to gradually send DATA frames, thereby effectively controlling the kept-open time of the CDN-website connection, exhaust all available connection resources of the host website, and ultimately launch a DoS attack that blocks requests from legitimate users.
\end{itemize}

We evaluated both of our attacks against six major CDN providers. Our results show that five of them are vulnerable to HCA attacks, and all six are vulnerable to HBA attacks. We responsibly disclosed our findings to all CDN providers. So far, two CDN vendors have acknowledged their vulnerability and awarded us bug bounties.  They have deployed our mitigations.
To further quantify the potential real-world impact of our attacks, we design an automated measurement framework to discover potentially vulnerable domains. We apply our framework to evaluate subdomains from the Tranco Top 1M domains, identifying 151,685 subdomains hosted by vulnerable CDNs, and finding that 42,330 subdomains are potentially vulnerable to our attacks.

\noindent \textbf{Our Contributions.} We make the following contributions in this paper.
\begin{itemize}
\item[$\bullet$] \textbf{Two new types of amplification attacks.} We introduce \textit{two new types of amplification attacks}, the HBA and the HCA attack. These attacks exploit the HTTP/3-to-HTTP/1.1 conversion, exhausting the victim's bandwidth and connections. 

\item[$\bullet$] \textbf{Real-world evaluations.} 
We examine the HBA and HCA attacks on six popular CDN vendors to assess their feasibility and real-world impacts. The experiment results show that all examined CDNs are vulnerable to the HBA attack, and five of them are vulnerable to the HCA attack. 

\item[$\bullet$] \textbf{Real-world impact.} 
We perform a large-scale measurement across the Tranco Top 1M domain list using our framework, discovering 151,685 subdomains hosted by vulnerable CDNs, and identifying 42,330 subdomains potentially vulnerable to our attacks.

\item[$\bullet$] \textbf{Mitigation and responsible disclosure.} We present the mitigations and responsibly disclose all vulnerabilities to CDN vendors. In particular, two CDN vendors acknowledged vulnerabilities, rewarded our reports, and have deployed our mitigations.

\end{itemize}

\section{Overview}\label{sec:overview}

In this section, we first introduce some background on Content Delivery Networks (CDNs) and the new HTTP/3 protocol. Then, we present our problem statement, the threat model, and techniques to bypass the CDN cache mechanism.

\subsection{Background: CDN and HTTP/3}

CDN is a group of servers located in different geographical locations that have become the backbone of the Internet. It accelerates content delivery through edge servers and offers DDoS protection by absorbing most of the attack traffic.
Specifically,
 CDN acts as a \textit{man-in-the-middle} between clients and websites to decouple traditional client-website connections into two segments: client-CDN and CDN-website connections. CDNs support different versions of the HTTP protocol in the two connections. 

HTTP/3 is a relatively new protocol proposed in 2022, which aims to reduce latency and optimize network performance.
The main innovation in HTTP/3 is the adoption of the QUIC (Quick UDP Internet Connections) protocol~\cite{rfc9000}, which distinguishes it from previous HTTP versions that run over TCP (Transmission Control Protocol)~\cite{rfc9293}. This fundamental shift is designed to solve the inherent ``head-of-line blocking'' issue that TCP faces in modern network environments, thereby further enhancing transmission efficiency and user experience. 
 Next, we will detail two optimization features supported by HTTP/3 that are relevant to our research.
\begin{itemize}
\item QPACK header compression~\cite{rfc9204}. QPACK is a header field compression format for HTTP/3 that makes the HPACK header compression format of HTTP/2 compatible with the QUIC protocol. Its primary purpose is to drastically shrink the size of HTTP headers that travel across the network. QPACK achieves this by referencing header fields that have been seen before, sending just small index numbers instead of the full header fields.
\item Multiplexing~\cite{rfc9114}. HTTP/3's multiplexing capability is achieved  by its underlying QUIC protocol. QUIC implements its own reliable transport, flow control, and multiplexing functions over UDP~\cite{rfc768}, operating on a connection-based multiplexing model. It allows for the simultaneous handling of multiple independent request streams over a single QUIC connection. Consequently, even if a packet in one stream is lost, it will not block other streams within the same connection, effectively solving TCP's head-of-line blocking problem.
\end{itemize}

\subsection{Problem Statement: Heterogeneous Deployment of HTTP/1.1 and HTTP/3}

Currently, all CDN providers claim to support HTTP/3~\cite{cloudflare-support-http/3} as shown in Table~\ref{tab: support http/3}.  The deployment is heterogeneous, though, which means that the client-CDN connection is HTTP/3, but the CDN-website connection is HTTP/1.1.  The reason is that many websites still do not support HTTP/3, making an end-to-end deployment infeasible.  Specifically, we conducted experiments on the HTTP/3 support of six popular CDN providers.
The experimental results, as shown in Table~\ref{tab: support http/3}, reveal that all CDNs support HTTP/3 in the client-CDN connection but only support HTTP/1.1 in the CDN-website connections, even when the host website supports HTTP/3.  Such a heterogeneous deployment leaves potential room for DoS attacks. 

\begin{table}[t]
\centering
\caption{CDN vendors' support for the HTTP/3 protocol.}
\resizebox{\linewidth}{!}{
\begin{tabular}{lccc}
\toprule[1.5pt]
           & client-CDN        & CDN-website & HTTP/3 Support           \\
\midrule
\midrule
Alibaba    & HTTP/1.1 \& HTTP/3 & HTTP/1.1   & Default Off Configurable \\
Baidu      & HTTP/1.1 \& HTTP/3 & HTTP/1.1   & Default Off Configurable \\
Cloudflare & HTTP/1.1 \& HTTP/3 & HTTP/1.1   & Default On  Configurable \\
CloudFront & HTTP/1.1 \& HTTP/3 & HTTP/1.1   & Default On Configurable  \\
Fastly     & HTTP/1.1 \& HTTP/3 & HTTP/1.1   & Default Off Configurable \\
Tencent    & HTTP/1.1 \& HTTP/3 & HTTP/1.1   & Default Off Configurable \\
\bottomrule[1.5pt]
\end{tabular}
}
\label{tab: support http/3}
\end{table}

\subsection{Threat Model}

Our threat model involves three parties in a CDN deployment.  
Specifically, we list each party and their capability below. 

\begin{itemize}

\item \textbf{An end web user:} a potential attacker. An attacker, as a web user, can craft malicious yet legitimate requests and send them to a CDN provider, which may then reroute the requests to host websites.  
\item \textbf{CDN Provider:} an unwitting intermediary. CDN providers act as a man-in-the-middle between end users and host websites. In our threat model, attackers exploit the CDN's HTTP/3-to-HTTP/1.1 conversion mechanism to amplify malicious traffic. The CDN itself is not the target of the attack but is abused as an amplifier to relay and magnify attack traffic toward the host website.
\item \textbf{Host Websites:} the victims. Host websites are the ultimate victims, which receive amplified DoS traffic from CDN providers, indirectly initiated by the malicious end user.
\end{itemize}

\subsection{Techniques to Bypass the CDN Cache Mechanism}
To send successive attack requests to the victim, the presence of a CDN cache may prevent forwarding of attack requests to the victim. When attacking a website, the victim may return a cacheable HTTP response. It is necessary to bypass the CDN cache mechanism to ensure that attacking requests reach the victim rather than hitting the CDN cache.
After exploring the CDN forwarding strategies and working mechanism exhaustively, we conclude several approaches to bypass the CDN cache mechanism and confirm them in the six CDN vendors, as listed in Table~\ref{table: bypass cache}.
CDN vendors cache resources based on HTTP-related parameters such as file extensions, URL paths, or HTTP Header. Attackers can exploit these rules to bypass CDN caching.

\begin{table}[t]
\centering
\caption{Techniques to bypass the CDN cache mechanism.}

\resizebox{\linewidth}{!}{
\begin{tabular}{lcccccc}
\toprule[1.5pt]
     & Alibaba & Baidu & Cloudflare & CloudFront & Fastly & Tencent \\
\midrule
\midrule
Dynamic Resources     & \CheckmarkBold              & \CheckmarkBold            & \CheckmarkBold                 & \CheckmarkBold                 & \CheckmarkBold             & \CheckmarkBold              \\
Random URL           & \CheckmarkBold              & \CheckmarkBold            & \CheckmarkBold                 & \CheckmarkBold                 & \CheckmarkBold             & \CheckmarkBold              \\
Query Parameters     & \CheckmarkBold              & \CheckmarkBold            & \CheckmarkBold                 & \CheckmarkBold                 & \CheckmarkBold             & \CheckmarkBold              \\
HTTP POST/PUT        & \CheckmarkBold              & \CheckmarkBold            & \CheckmarkBold                 & \CheckmarkBold                 & \CheckmarkBold             & \CheckmarkBold              \\
\makecell{WebSocket handshake}  & \CheckmarkBold              & \CheckmarkBold            & \CheckmarkBold                 & \CheckmarkBold                 & \CheckmarkBold             &                             \\
Cookie Header        & \CheckmarkBold              & \CheckmarkBold            & \CheckmarkBold                 &                                & \CheckmarkBold             &                             \\
\makecell{Authorization Header} & \CheckmarkBold              & \CheckmarkBold            & \CheckmarkBold                 &                                & \CheckmarkBold             &                             \\
\bottomrule[1.5pt]
\end{tabular}
}
\label{table: bypass cache}
\end{table}

\section{Design}\label{sec: H3 attack}

In this section, we first present the core design of our HTTP/3 amplification attack and then describe a measurement framework to quantify the impacts of the attacks for real-world websites. 

\subsection{HTTP/3 Amplification Attack}

\begin{figure}[!t]
    \centering
    \includegraphics[width=0.47\textwidth]{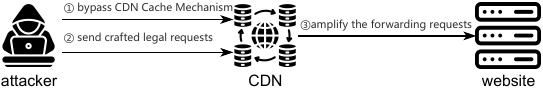}
    \caption{Attack concept of HTTP/3 Amplification Attacks.}
    \label{fig: attack concept}
\end{figure}

Figure~\ref{fig: attack concept} shows the overall attack flow of our amplification attack. The high-level idea is to amplify the connection from an attacker to a CDN to a huge request (either in terms of request size or the number of requests) from the CDN to the host website.  Specifically, the attack has three steps: (1) the attacker bypasses the CDN cache mechanism; (2) the attacker sends crafted legal requests; and (3) the CDN amplifies the forwarding requests.

The first step is to bypass the CDN cache mechanism. As discussed in Section~\ref{sec:overview}, we summarize several techniques to bypass the CDN cache mechanism (Table~\ref{table: bypass cache}). In this paper, we primarily use HTTP POST requests to bypass the CDN cache. The next steps are launching attacks, and we describe two types of such amplifications below.

\subsubsection{\textbf{Bandwidth Amplification}}\label{sec: HBA attack}

\begin{figure}[!t]
    \centering
    \includegraphics[width=0.47\textwidth]{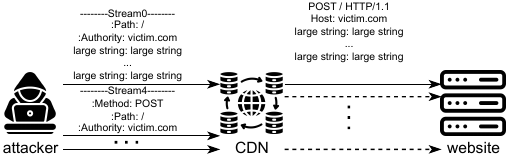}
    \caption{Concept of HTTP/3 Bandwidth Amplification Attack. An attacker sends HTTP/3 requests with QPACK headers; the CDN must decompress the HTTP/3 headers, resulting in bandwidth amplification.}
    \label{fig: HBA}
\end{figure}

The HTTP/3 Bandwidth Amplification (HBA) attack enables attackers to exploit a CDN to convert small HTTP/3 requests into significantly larger HTTP/1.1 requests, thereby depleting the target host website's bandwidth.
Figure~\ref{fig: HBA} shows the attack principle of our HBA attack. Since HTTP/1.1 lacks this header compression mechanism, the CDN forces all indexed fields to be expanded into large raw HTTP/1.1 headers, which leads to a significant bandwidth amplification attack.

There are two variations of the HBA attack, depending on whether static and/or dynamic tables are supported. Specifically, according to the QPACK mechanism, in the client-CDN connection, both the client and the CDN maintain static and dynamic tables. The static table is predefined and includes a set of the most common HTTP header fields and their values. Both the client and CDN implicitly possess the same static table, allowing direct referencing via indexes during transmission, eliminating the need to transmit the full content of these common fields. The dynamic table is updated dynamically based on the actual header fields transmitted after the connection is established. The client adds new header fields appearing in the connection to the dynamic table and assigns them indexes. Subsequent duplicate header fields are replaced by referencing the corresponding index values in the dynamic table via their indexes, significantly reducing the number of bytes transmitted. Although QPACK can bring significant performance improvements, its support varies among CDNs.

If dynamic tables are supported, an attacker can initiate the attack by first sending an HTTP/3 request with a large header that is inserted into the dynamic table. For subsequent requests, the attacker then repeatedly utilizes small QPACK index values to reference this raw large header.  Currently, static tables are widely supported, but only half of the existing CDNs support dynamic tables, as shown in Table~\ref{table: amplification factor of static table}.

\subsubsection{\textbf{Connection Amplification}}\label{sec: HCA attack}

The Connection Amplification enables attackers to manipulate the kept-open time of CDN-website connections to exhaust all available connection resources of the host website.
 Specifically, to accelerate content delivery, CDNs establish CDN-website connections immediately upon receiving HTTP/3 HEADERS frames.
  Therefore, an attacker can exploit this behavior by slowly sending HTTP/3 DATA frames to indirectly control the kept-open time of CDN-website connections, thereby launching an HCA attack.

\begin{figure}[t]
    \centering
    \includegraphics[width=0.47\textwidth]{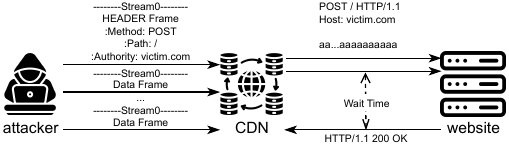}
    \caption{Concept of HTTP/3 Connection Amplification Attack. An attacker sends HTTP/3 HEADERS frames in an HTTP/3 connection; the CDN must establish multiple HTTP/1.1 connections to the host website, resulting in connection amplification.}
    \label{fig: HCA}
\end{figure}

Figure~\ref{fig: HCA} shows the attack principle.
 An attacker first sends a HEADERS frame in an HTTP/3 stream to force the CDN to establish an HTTP/1.1 connection with the host website. Subsequently, the attacker prompts the CDN to maintain the CDN-website connection for an extended period by sending subsequent DATA frames at a very low rate. This attack leads the CDN to believe that the request is not yet complete, thus continuously occupying the host website's connection resources. Since most CDNs support HTTP/3 in the client-CDN connection but still only support HTTP/1.1 in the CDN-website connection, this protocol conversion mechanism means that the CDN will establish an individual TCP connection with the host website for each independent HTTP/3 request stream.

 HTTP/3's multiplexing feature allows attackers to initiate a large number of concurrent HTTP/3 request streams within a single HTTP/3 connection. Specifically, the HTTP/3 protocol supports the simultaneous processing of multiple bidirectional streams within a single HTTP/3 connection, thereby avoiding head-of-line blocking during the TCP handshake and achieving full multiplexing of requests and responses.

  Each HCA attack stream will prompt the CDN to establish a separate connection between the CDN and the host website. Consequently, with minimal resource consumption on their part, an attacker can force CDNs to establish a large number of CDN-website connections far exceeding their own needs, which significantly enhances the attack's efficiency and amplification effect. Ultimately, all available connection resources of the host website will be exhausted, preventing it from responding to legitimate user requests, and thereby achieving a DoS attack.

\subsection{Attack Impact Measurement}
\begin{figure*}[!t]
    \centering
    \includegraphics[width=0.8\linewidth]{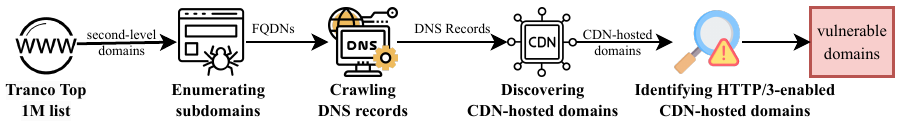}
    \caption{Workflow of the automated measurement framework.}
    \label{fig: workflow}
\end{figure*}

Quantifying the potential real-world impact of HTTP/3 amplification attacks is crucial. To achieve this, we designed and implemented an automated measurement framework that can identify websites simultaneously hosted on vulnerable CDNs with HTTP/3 support enabled. 
The workflow of our automated measurement framework is divided into four main steps, as shown in Figure~\ref{fig: workflow}.

\noindent\textbf{Enumerating subdomains.} The objective of this step is to acquire a collection of Fully Qualified Domain Names (FQDNs) from the Tranco Top 1M list. We employ a brute-force technique based on a custom prefix dictionary. First, we generate a list of FQDNs under target second-level domains (SLDs) using this custom prefix dictionary, providing greater flexibility in the discovery process. We append each prefix to the target SLD to create an FQDN and verify its existence through DNS queries. We utilize this method to enumerate FQDNs for the Tranco Top 1M domain list.

\noindent\textbf{Crawling DNS records.} After obtaining the FQDNs list, we crawl the CNAME and NS records for each FQDN. To speed up the DNS record crawling for the Tranco Top 1M domains, we launched five DNS crawler instances simultaneously. Each instance began crawling from a different index within the Tranco Top 1M list.

\noindent\textbf{Discovering CDN-hosted domains.} When a domain owner wishes to deploy their domains to a CDN, the CDN typically assigns a subdomain belonging to the CDN. The domain owner must then create a CNAME record, pointing their domain to this CDN-assigned subdomain to redirect traffic to the CDN. We identify CDN-hosted domains through the following steps: First, for a given CDN (e.g., Fastly), we collect CDN-assigned subdomains from previous work~\cite{lin2024detecting} (Table~\ref{table: cdn_assigned_subdomain}). Second, we compare the DNS records with these CDN-assigned subdomains to identify vulnerable CDN-hosted domains.

\begin{table}[t]
    \caption{CDN-assigned subdomains.}
    \centering
    \begin{threeparttable}
    \begin{tabular}{lc}
    \toprule[1.5pt]
    CDN            & \textbf{Assigned subdomain}                       \\
    \midrule
Alibaba        & .cdngslb.com, .alicdn.com, .alikunlun.com           \\
Baidu          & .bdydns.com, .jomodns.com, .yunjiasu-cdn.net        \\
Cloudflare     & .cdn.cloudflare.net                                 \\
CloudFront     & .cloudfront.net                                     \\
Fastly         & .fastly.net, .fastlylb.net                          \\
Tencent        & .cdn.dnsv1.com, .tdnsv6.com                         \\
    \bottomrule[1.5pt]
    \end{tabular}
    \end{threeparttable}
    \label{table: cdn_assigned_subdomain}
\end{table}

\noindent\textbf{Identifying HTTP/3-enabled CDN-hosted domains.} We utilize aioquic as a client to send HTTP/3 requests to vulnerable CDN-hosted domains. If the server responded successfully, it indicated that the host website had HTTP/3 support enabled on that CDN. We consider these as vulnerable domains.
\section{Evaluation}\label{sec: real-world}

Our evaluation answers the following research questions. 

\begin{itemize}

\item \textit{\textbf{RQ1}: How many real-world CDNs are vulnerable to HTTP/3 amplification attacks?}

{\textbf{Answer}: We evaluated both of our attacks against six major CDN providers in the real world. Our results show that all six of them are vulnerable to HBA attacks, and five of them are vulnerable to HCA attacks.}

\item \textit{\textbf{RQ2}: How many websites are affected by the proposed HTTP/3 amplification attacks?}

{\textbf{Answer}: We apply our framework to evaluate subdomains from the Tranco Top 1M domain list~\cite{tranco}, identifying 151,685 subdomains hosted by vulnerable CDNs, and finding that 42,330 subdomains are potentially vulnerable to our attacks.}

\item \textit{\textbf{RQ3}: Which CDNs immediately establish the CDN-website connection upon receiving HTTP/3 HEADERS, and what is the maximum connection kept-open time?}

{\textbf{Answer}: Five CDNs immediately establish the CDN-website connection upon receiving HTTP/3 HEADERS frames, and their kept-open time depends on the client-CDN connection's kept-open time, which makes them vulnerable to the HCA attack.}

\item \textit{\textbf{RQ4}: How does the number of concurrent streams affect this amplification factor?}

{\textbf{Answer}: When the attacker uses the static table, the amplification factor does not change with the number of concurrent streams; however, when using the dynamic table, the amplification factor increases with the number of concurrent streams, peaking at approximately 64 streams with a maximum amplification factor of approximately 350$\times$ due to the CDN's CPU overhead in decompressing QPACK indexes into large HTTP/1.1 headers.}

\end{itemize}

\subsection{RQ1: Real-world Vulnerable CDNs}

In this research question, we evaluate both HBA and HCA attacks against six real-world CDNs.

\noindent \textbf{Experiment Setup.} We set up an Apache HTTP server in Hong Kong as the target host website, with the Apache \texttt{TimeOut} value of 300 seconds and a connection limit of 256. The attacker was launched from a VPS in Singapore. For ethical considerations, we deliberately limited the victim website's bandwidth to 100Mbps and the attacker's bandwidth to 30Mbps to ensure that our experimental traffic would not degrade CDN network performance or impact other CDN-hosted websites. It is important to note that these bandwidth constraints represent our self-imposed ethical limits, not the upper bounds of the attack's actual capability. In real-world scenarios, websites with higher bandwidth and connection limits would require proportionally more attack streams, but the fundamental attack mechanism remains effective, as the amplification ratio is inherent to the HTTP/3-to-HTTP/1.1 conversion and is independent of the victim's capacity.

\subsubsection{RQ1.1: HBA Attacks.}
\begin{figure*}[t]
	\subfigure[Bandwidth consumption of the attacker] %
	{
		\begin{minipage}{8.5cm}
			\centering          
			\includegraphics[scale=0.28]{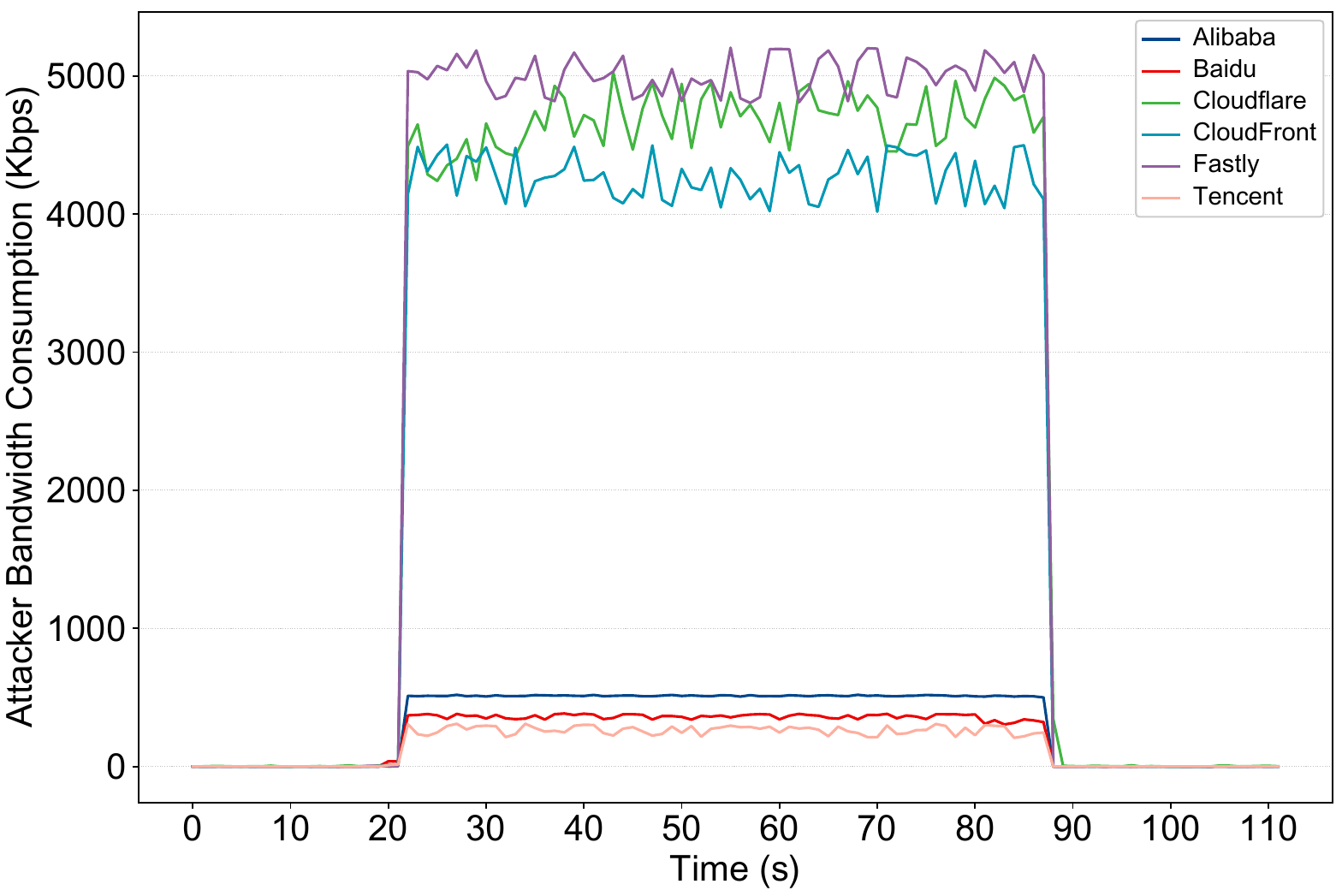}   
		\end{minipage}
            \label{fig:hba attacker bandwidth}
	}
	\subfigure[Bandwidth consumption of the host website] 
	{
		\begin{minipage}{8.5cm}
			\centering      
			\includegraphics[scale=0.28]{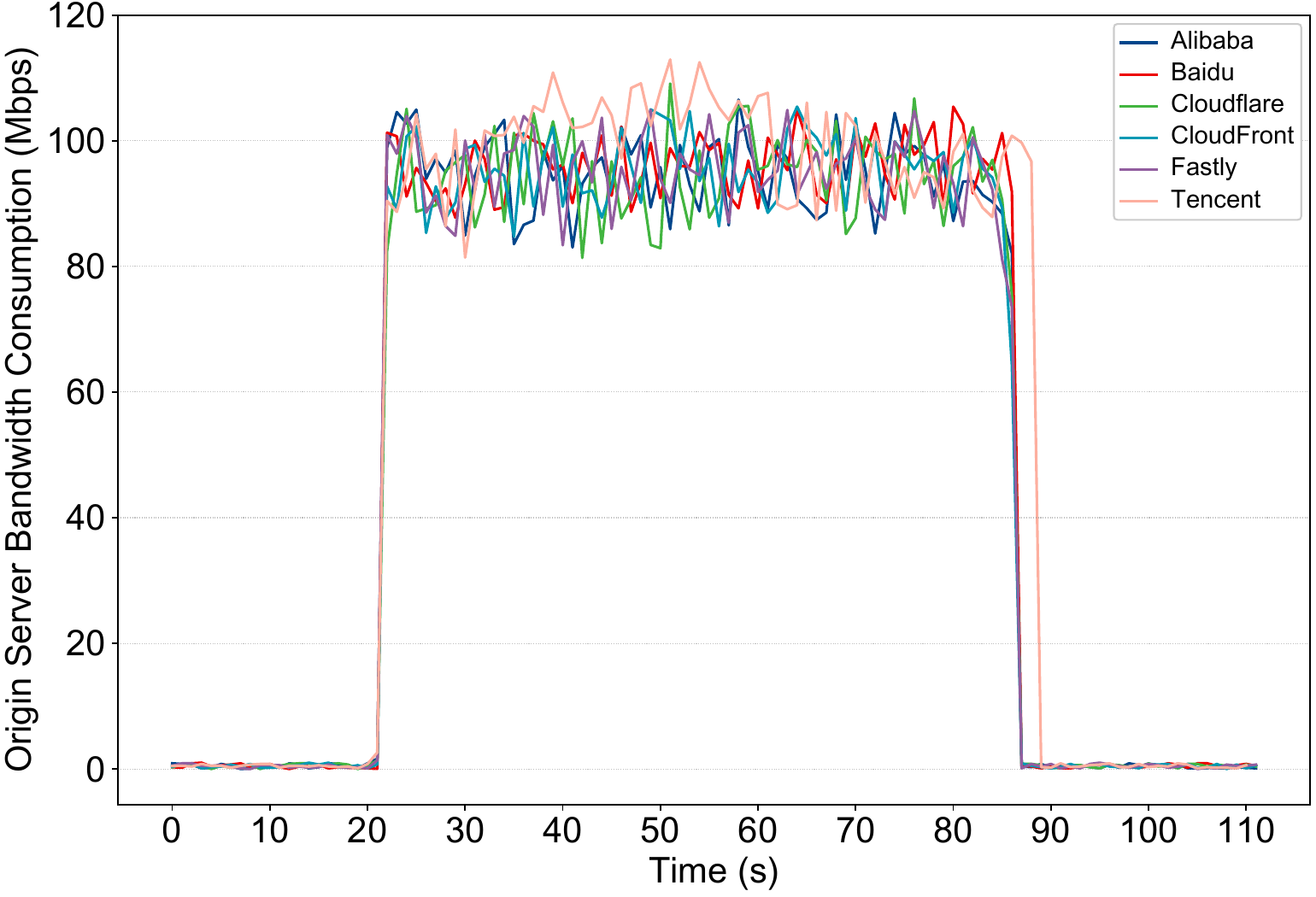}   
		\end{minipage}
            \label{fig:hba origin bandwidth}
	}
        \caption{Launch an HBA attack lasting 60s.}
	\label{fig:hba attack}   
\end{figure*}

Figure~\ref{fig:hba origin bandwidth} shows the experiment results. For each vulnerable CDN, the bandwidth of the host website is consistently greater than 100Mbps during the HBA attack. Meanwhile, as shown in Figure~\ref{fig:hba attacker bandwidth}, the bandwidth consumption of attackers from the three vendors that support dynamic tables is all less than 500Kbps, while that of attackers from the other vendors that only support static tables is all less than 5Mbps. These results demonstrate that the DoS attack was successfully carried out, depleting the bandwidth resources of host websites across different CDNs.

\begin{figure}[t]
    \centering
    \includegraphics[width=0.4\textwidth]{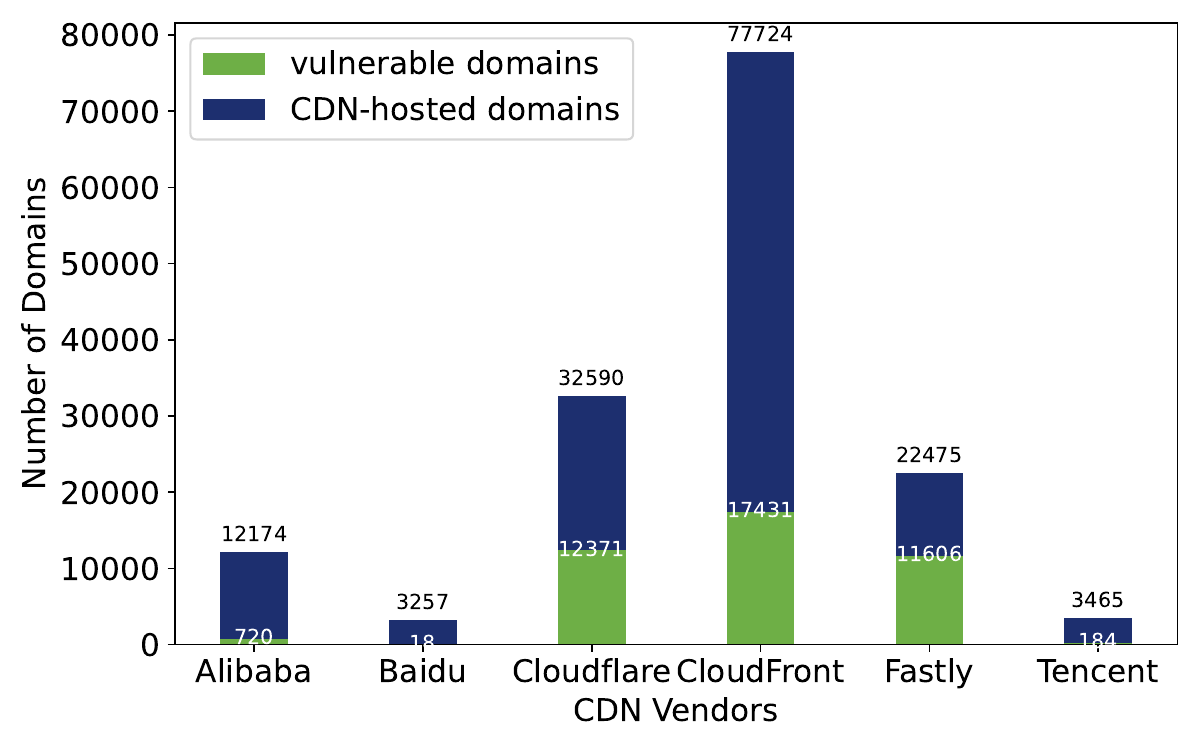}
    \caption{Number of potentially vulnerable domains across different CDN providers in the Tranco Top 1M list.}
    \label{fig:measurement_results}
\end{figure}

\subsubsection{RQ1.2: HCA Attacks}

Table~\ref{table: amplification factor of static table} shows that all six CDNs support a minimum of 100 concurrent streams. For most CDNs, by establishing four HTTP/3 client-CDN connections and multiplexing 96 streams within each connection ($4 \times 96 = 384$), we can force each CDN to establish 384 HTTP/1.1 CDN-website connections, which effectively exhausts all available connections on the host website. Fastly establishes a maximum of 10 CDN-website connections per HTTP/3 connection. Therefore, we initiated 48 HTTP/3 client-CDN connections, each multiplexing 8 streams ($48 \times 8 = 384$), which also establish 384 HTTP/1.1 CDN-website connections on the host website.
On the attacker side, we first send HTTP request headers for each stream using the HTTP/3 HEADERS frame. Subsequently, we slowly send the HTTP request body via the HTTP/3 DATA frame, which allows us to maintain the CDN-website connection and carry out a continuous HCA attack for 300 seconds. During the HCA attack, we periodically send GET requests to the CDN every second and record the CDN response time to probe whether the connection resources of the host website have been exhausted.

\begin{figure*}[t]	
	\subfigure[Number of established CDN-website connections] %
	{
		\begin{minipage}{8.5cm}
			\centering          
			\includegraphics[scale=0.28]{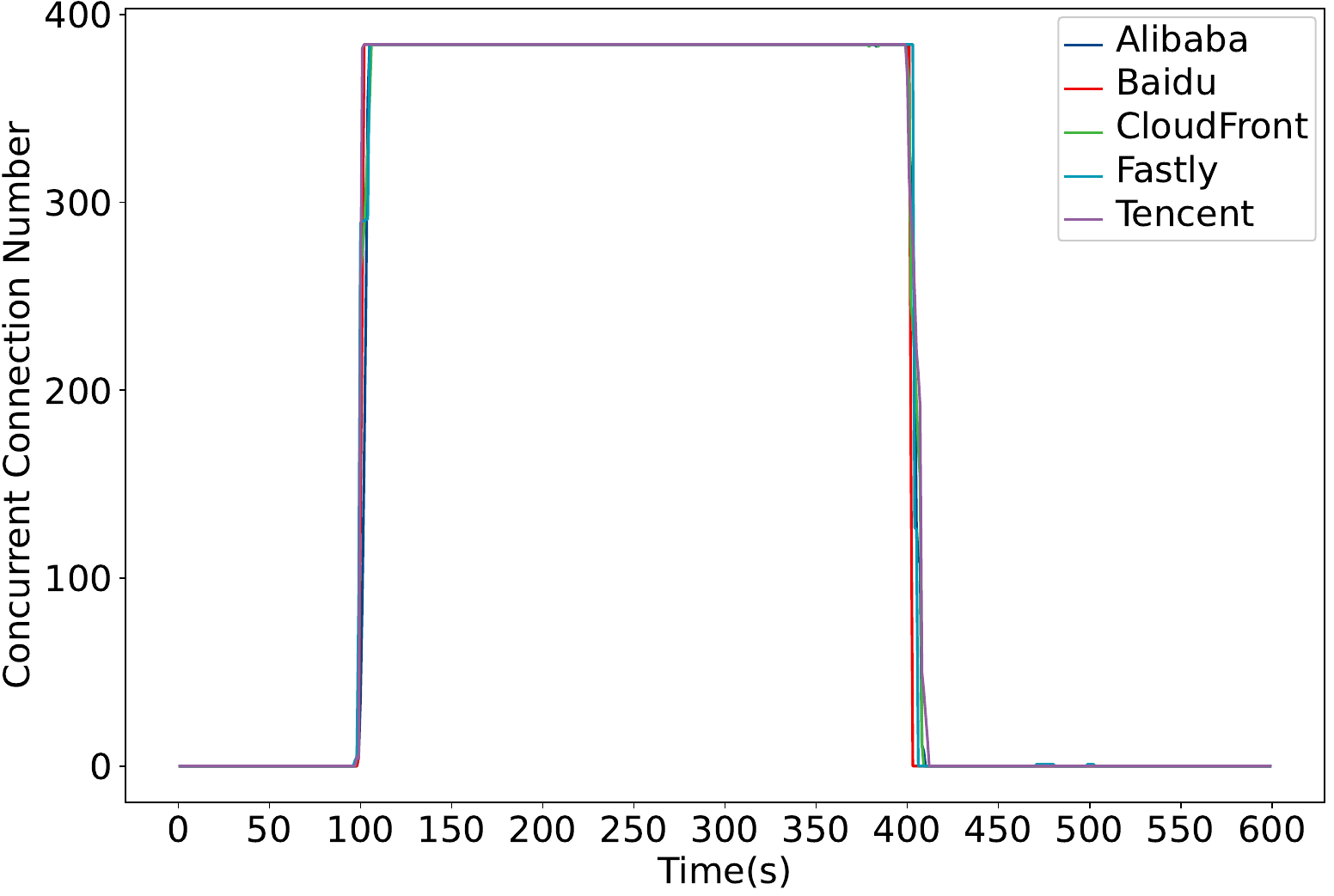}   
		\end{minipage}
            \label{fig:hca number of connections}
	}
	\subfigure[CDNs response time for the benign client's request] 
	{
		\begin{minipage}{8.5cm}
			\centering      
			\includegraphics[scale=0.28]{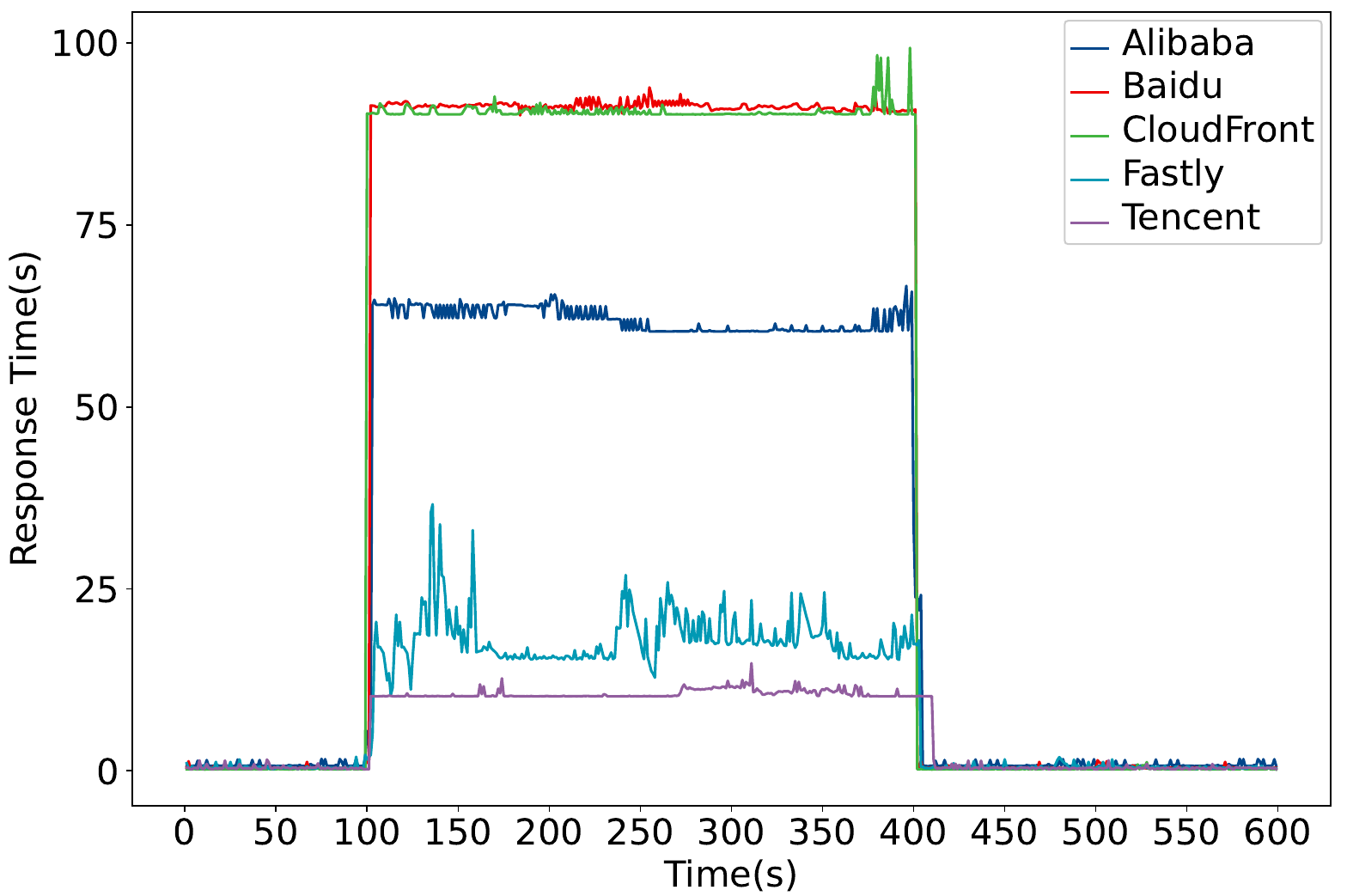}   
		\end{minipage}
            \label{fig:hca response time}
	}
        \caption{Launch an HCA attack lasting 300s.}
	\label{fig:hca attack}   
\end{figure*}

Figure~\ref{fig:hca number of connections}
 shows the experiment results of our HCA attacks.  For each vulnerable CDN, the number of established CDN-website connections is consistently greater than 300 during the HCA attack. At the host website, connection resources are exhausted, and other requests are starving. Consequently, as shown in Figure~\ref{fig:hca response time}, when we launch the HCA attack at 100s, the response time for benign clients reached 60s on Alibaba and up to 90s on Baidu and CloudFront, returning an HTTP 504 Gateway Timeout. In particular, for Tencent, approximately 10 seconds after sending a GET request to probe the host website, it actively closes the client-CDN connection, resulting in no response to the probe request. Meanwhile, Fastly's response time rises to 15s, returning an HTTP 503 Service Unavailable. These results demonstrate that the DoS attack was successfully executed, severely affecting the availability of host website services across different CDNs.

\subsection{RQ2: Measurement of Impacted Websites}

The purpose of this research question is to measure how many real-world websites are impacted by those two attacks. 
 Specifically, our large-scale measurement across the Tranco Top 1M domain list provides a quantitative assessment of the real-world attack surface for our attacks. Figure~\ref{fig:measurement_results} illustrates the distribution of CDN-hosted domains and, crucially, the subset of ``potentially vulnerable domains'' (i.e., those hosted on identified vulnerable CDNs and having HTTP/3 support enabled, thus satisfying all prerequisites for our attacks) across six prominent CDN providers.
As illustrated in Figure~\ref{fig:measurement_results}, 151,685 domains were hosted by vulnerable CDNs. Of these, 42,330 are identified as potentially vulnerable. CloudFront has the highest number of domains among vulnerable CDNs, hosting a total of 77,724 domains. Of these, 17,431 are identified as potentially vulnerable.
Cloudflare ranks second with 32,590 domains, of which 12,371 are potentially vulnerable. Fastly reveals a total of 22,475 domains, with 11,606 identified as potentially vulnerable. Alibaba hosts 12,174 domains, of which 720 are potentially vulnerable, while Tencent has 3,465 domains, with 184 classified as potentially vulnerable. Baidu has the fewest domains, hosting only 3,257, of which 18 are potentially vulnerable to our attacks.
These results underscore the widespread presence of potentially vulnerable websites within the current Internet ecosystem, particularly across major CDN services.

\subsection{RQ3: Kept-Open Time of CDN-website Connection}

The purpose of this research question is to explore how the kept-open time of CDN-website connection affects our HCA attack.

\noindent \textbf{Experiment Setup.} In our experiments, we set up a self-hosted Apache server and deployed it as the host website behind six CDNs, one at a time. For each CDN, we first send HTTP/3 HEADERS frames, and then send a DATA frame per second, taking 300 seconds to complete the transmission. 
We establish an HTTP/3 connection with the CDN and send 100 concurrent requests within this connection. Concurrently, we use tcpdump on the host website to capture the time it took for the CDN to establish a CDN-website connection, as well as the kept-open time of CDN-website connection.

\noindent \textbf{Experiment Results.} 
After sending 100 concurrent requests and repeating this process 10 times, we obtain the average results as shown in Table~\ref{table: connection time}. We can observe that Alibaba, Baidu, CloudFront, Fastly, and Tencent immediately begin establishing a CDN-website connection upon receiving the HEADERS frame. In contrast, Cloudflare only established the connection after receiving the complete HTTP/3 request. Clearly, for Alibaba, Baidu, CloudFront, and Tencent, the kept-open time of the CDN-website connection depends on the kept-open time of the client-CDN connection, which is directly controlled by the client, thus potentially exploitable by malicious attackers. Although Fastly also immediately began establishing a CDN-website connection upon receiving the HEADERS frame, its behavior differs: it establishes CDN-website connections for the first 10 HTTP/3 HEADERS frames within each HTTP/3 connection. Subsequent requests within the same HTTP/3 connection are then queued, during which Fastly must wait to receive the complete requests, resulting in a kept-open time of 5.46s.

\begin{table}[t]
\centering
\caption{Connection establishing and kept-open times when sending 100 streams for different CDN vendors.}
\resizebox{\linewidth}{!}{
\begin{tabular}{lcccccc}

\toprule[1.5pt]
                          & Alibaba & Baidu  & Cloudflare & CloudFront & Fastly & Tencent \\
\midrule
\midrule

\makecell{Connection\\establishing Time}    & 2.88  & 1.80  & 301.56 & 2.94    & \makecell{1.33(10)\\304.97(90)}& 2.66   \\
\makecell{Connection\\Kept-open Time} & 311.97 & 307.76 & 7.57    & 306.53 & \makecell{308.84(10)\\5.46(90)}& 309.96 \\
\bottomrule[1.5pt]
\end{tabular}
}
\label{table: connection time}
\end{table}

\subsection{RQ4: Amplification Factor Analysis}

The purpose of this research question is to perform an amplification factor analysis for our HBA attack. 

\noindent \textbf{Experiment Setup.} 
We performed the HBA attack using two distinct QPACK compression strategies to determine the maximum amplification factor. We initially utilized the standard QPACK static table, and subsequently leveraged the QPACK dynamic table. 

\noindent \textbf{Experiment Results.} 
Both the client and the server possess identical static tables, allowing for direct referencing via indexes during transmission. Consequently, in an HBA attack that leverages the static table, the amplification factor remains independent of the number of streams. In our experiments, we utilize index 56 from the static table to achieve the amplification factors shown in Table~\ref{table: amplification factor of static table}.
\begin{table}[t]
\caption{Maximum amplification factors across the CDNs using static table.}
\resizebox{\linewidth}{!}{
\begin{threeparttable}
\begin{tabular}{lcccccc}
\toprule[1.5pt]
                       & Alibaba & Baidu & Cloudflare & CloudFront & Fastly & Tencent \\
\midrule
\midrule
\# Streams              & 128     & 128   & 256        & 128        & 100    & 128     \\
\# Headers & 950     & 350   & 190        & 330        & 210    & 201     \\
Factor   & 65.8    & 66.06 & 48.27      & 51.2       & 36.41  & 54.08  \\
\makecell{Dynamic Table} & \CheckmarkBold              & \CheckmarkBold            & \XSolidBrush                    & \XSolidBrush                    & \XSolidBrush                & \CheckmarkBold             \\
\bottomrule[1.5pt] 
\end{tabular}
\begin{tablenotes}
\item[\raisebox{-1.0ex}{\CheckmarkBold}] This CDN supports the QPACK dynamic table.
\item[\raisebox{-1.0ex}{\XSolidBrush}] This CDN does not support the QPACK dynamic table.
\end{tablenotes}
\end{threeparttable}
\label{table: amplification factor of static table}
}
\end{table}

In contrast, for an HBA attack that leverages the dynamic table, the amplification factor is influenced by the number of concurrent streams. The maximum value for concurrent streams is negotiated during the establishment of the HTTP/3 connection. Table~\ref{tab: limits} summarizes the limits set by each CDN on HTTP/3 streams and dynamic tables. All three CDNs that support dynamic tables allow a maximum of 128 concurrent streams, with a dynamic table size of 4KB and a maximum entry size of 3,072 bytes.

\begin{table}[t]
\caption{Limits set by CDNs on HTTP/3 streams and dynamic tables.}
\resizebox{\linewidth}{!}{
\begin{tabular}{lcccccc}
\toprule[1.5pt]
              & Alibaba & Baidu & Cloudflare & CloudFront & Fastly & Tencent \\
              \midrule
              \midrule
Max Streams & 128     & 128      & 256        & 128        & 100    & 128     \\
Dynamic Table Size    & 4KB     & 4KB   & N/A        & N/A        & N/A    & 4KB     \\
Max Entry Size        & 3072B   & 3072B & N/A      & N/A      & N/A  & 3072B  \\
\bottomrule[1.5pt]
\end{tabular}
}
\label{tab: limits}
\end{table}

\begin{figure}[t]
    \centering
    \includegraphics[width=0.4\textwidth]{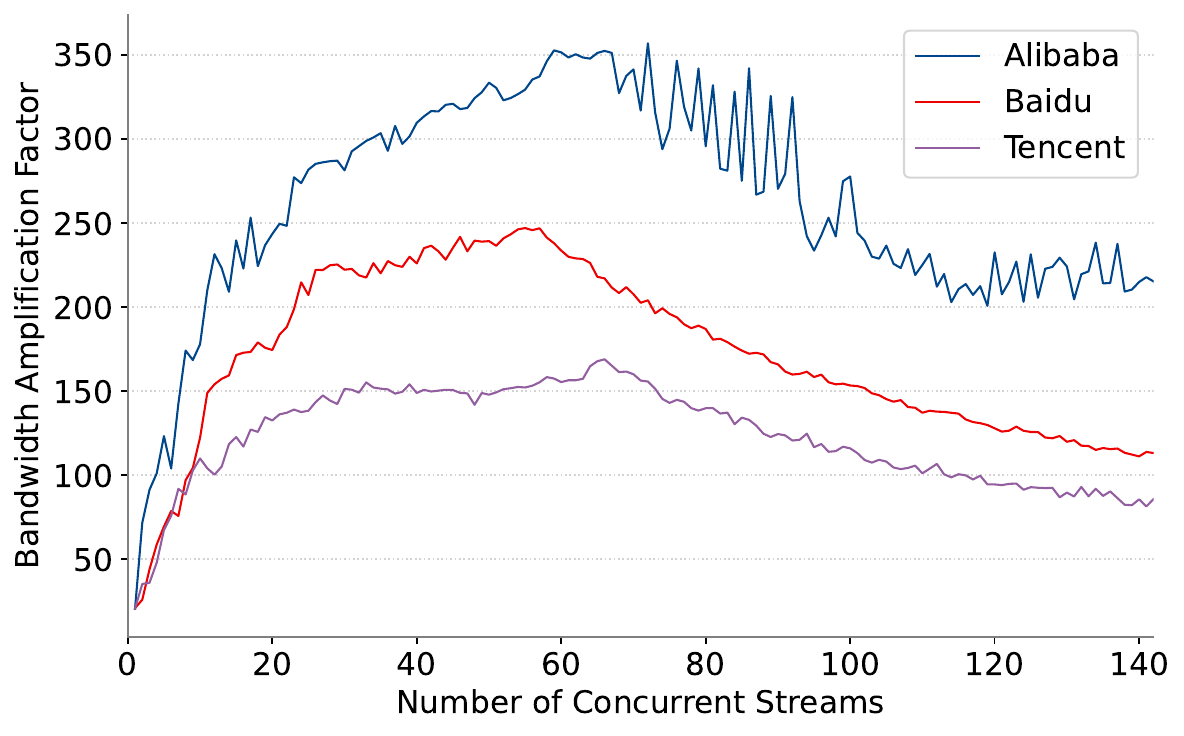}
    \caption{Bandwidth amplification factor when the number of concurrent streams increases.}
    \label{fig: streams amplification}
\end{figure}

To explore how the number of concurrent streams affects the amplification factor, we experimented using a single HTTP/3 connection, varying the number of streams from 1 to 140. We capture traffic on both the client and the host website using tcpdump to evaluate the amplification factor. As shown in Figure~\ref{fig: streams amplification}, the bandwidth amplification factor increases with the number of concurrent streams. The amplification factor peaks at approximately 64 concurrent streams rather than at the maximum of 128, reaching a maximum bandwidth amplification factor of approximately 350$\times$. This is because the CDN must decompress each QPACK-indexed header into its full HTTP/1.1 representation before forwarding, which introduces significant CPU overhead at the CDN edge. When the number of concurrent streams increases beyond this point, the CDN's processing capacity becomes the bottleneck: subsequent streams must wait for previous streams to complete their header decompression and forwarding, effectively serializing the conversion process and preventing further amplification gains. When the number of streams exceeds the maximum allowed (128 streams), our HTTP/3 client must additionally wait for previous streams to close before opening new ones, causing the bandwidth amplification factor to fluctuate.

\section{Discussion}\label{sec:discussion}

\subsection{Severity Assessment}

\noindent \textbf{A severe and widespread practical influence.} According to our experimental results, the HBA and HCA attacks can effectively exhaust the host website's bandwidth and connection resources, respectively. As detailed in Section~\ref{sec: real-world}, the CDNs we tested enjoy global popularity and a high market share. These CDNs host a significant portion of websites on the Tranco Top 1M list: our measurement identifies 151,685 subdomains hosted by vulnerable CDNs, of which 42,330 have HTTP/3 enabled and satisfy all prerequisites for our attacks, making them potentially vulnerable. Therefore, many well-known websites are exposed to the proposed attacks.

\noindent \textbf{A low-cost and efficient DoS attack.} Unlike other DDoS attacks that need to control a large scale of botnets~\cite{Botnet-DDoS}, the attacker only needs an ordinary laptop to launch the HBA and HCA attacks. As the CDN nodes are dispersed globally, they form a naturally distributed amplification infrastructure~\cite{Mirai_Botnet, Botnet_Detection}, allowing attackers to easily congest the target network and possibly cause a denial of service in seconds for a negligible cost.

\noindent \textbf{Scalability to high-capacity host websites.} Our experiments were conducted against a 100Mbps host website due to ethical constraints. However, the amplification ratio is inherent to the HTTP/3-to-HTTP/1.1 conversion mechanism and is independent of the victim's capacity. The attack scales linearly: since our HBA attack with dynamic tables achieves an amplification factor of approximately 350$\times$, an attacker consuming less than 500Kbps can saturate a 100Mbps host website. Similarly, for the HCA attack, exhausting the connection resources of a higher-capacity host website with more concurrent connection slots simply requires proportionally more HTTP/3 streams, which can be trivially achieved by opening additional HTTP/3 connections. Therefore, our attacks remain practical and effective against well-provisioned production servers.

\noindent \textbf{A security challenge to anti-DDoS.} Traditional DDoS attacks primarily target the victim's incoming bandwidth through volumetric flooding. Our attacks, however, exploit legitimate HTTP/3 requests that pass through CDN infrastructure, making them significantly harder to detect and filter. Since the CDN itself is being abused to amplify the attack traffic, the malicious requests appear as normal CDN-to-host-website traffic, posing security challenges for existing DDoS detection mechanisms~\cite{DDoS_Detection, DoS_survey}.

\subsection{Ethics and Responsible Disclosure}

In our research, our goal was to strike a balance between assessing the severity of real-world attacks and minimizing the potential impact on CDN vendors. We were concerned that using excessive bandwidth during evaluation could degrade the CDN network performance and impact other CDN-hosted websites. Therefore, throughout our evaluations, we paid meticulous attention to avoiding any potential ethical issues. Firstly, we strategically chose attacker locations to ensure minimal latency to CDN servers, which allowed attacks to launch within the backbone network. This setup significantly reduced the risk of attack traffic saturating on-path network segments. Secondly, our attacks targeted our own websites, and we strictly controlled the traffic to remain well below the CDN's overall capacity, thereby preventing any disruption to other websites hosted on the CDN. Thirdly, we limited the bandwidth of the victim website to 100Mbps to further protect websites hosted on the CDN. These prudent methods ensured that our research effectively enhanced CDN security while minimizing any potential harm.

We responsibly contacted six CDN vendors to report all vulnerabilities found in our study, providing detailed reports and mitigations. All vendors acknowledged our disclosure and expressed appreciation. Two vendors, Baidu and Tencent, confirmed the vulnerabilities, rewarded our reports through their bug bounty programs, and have deployed our mitigations. The other vendors expressed gratitude and are still discussing the reported vulnerabilities internally. The responses from CDN vendors are summarized below:

\noindent \textbf{Baidu:} acknowledged our report, rated the HCA and HBA vulnerabilities as medium-severity and high-severity, with a bug bounty reward of approximately \$350, and deployed our mitigations.

\noindent \textbf{Tencent:} acknowledged our report, rated the HCA vulnerability as medium severity with a bug bounty reward of approximately \$150. They deployed our mitigations, which limit the number of CDN-website connections and restrict the size of HTTP headers in the dynamic table.

\subsection{Root Cause}
The root cause of HBA and HCA attacks lies in the fact that CDNs do not support end-to-end HTTP/3. CDNs support HTTP/3 in client-CDN connections, but only HTTP/1.1 for CDN-website connections, even if the host website supports HTTP/3. As a result, when forwarding requests, CDNs must convert HTTP/3 requests to HTTP/1.1. HTTP/3 offers significant efficiency through features like QPACK header compression and multiplexing, but these benefits are lost during this conversion. Worse still, an attacker can then exploit the HTTP/3-to-HTTP/1.1 conversion and HTTP/3 features to launch amplification attacks.

\subsection{Mitigation}
Both HBA and HCA attacks stem from the HTTP/3-to-HTTP/1.1 conversion behavior exhibited by CDNs. This conversion persists because CDN providers see a low return on investment for supporting HTTP/3 in CDN-website connections.
More critically, current HTTP/3 specifications~\cite{rfc9114} do not provide clear guidance on securing the HTTP/3-to-HTTP/1.1 conversion behavior.
Based on our analysis, we derive the following design principles and concrete mitigations for improving the reliability of CDN services under heterogeneous protocol deployments.

\noindent \textbf{Principle 1: Bound the amplification ratio at the protocol conversion layer.}
The HBA attack exploits the CDN's conversion of compact QPACK-indexed HTTP/3 headers into verbose HTTP/1.1 headers, resulting in significant bandwidth amplification. To mitigate this, CDNs should enforce limits at three levels:
\noindent (1) \textit{Limit the size of individual HTTP headers in the dynamic table.} CDNs should cap the maximum size of any single header field entry that can be inserted into the QPACK dynamic table (e.g., 512B per entry, about the size of a typical cookie). This prevents attackers from inserting extremely large headers that are later referenced by small index values. Tencent has deployed this mitigation following our disclosure.
\noindent (2) \textit{Restrict the number of times the same header index can be referenced within a single stream.} An attacker can reference the same large header hundreds of times within one stream to maximize the amplification factor. CDNs should impose a per-stream reference limit (e.g., no more than 10 references to the same dynamic table entry per stream). This directly caps the amplification ratio regardless of header size.
\noindent (3) \textit{Enforce a maximum decompressed request size.} CDNs should set an upper bound on the total decompressed HTTP/1.1 request size (e.g., 64KB). Any request exceeding this threshold after QPACK decompression should be rejected before forwarding to the host website.

\noindent \textbf{Principle 2: Validate complete requests before establishing backend connections.}
The HCA attack exploits the CDN's eager connection establishment behavior: most CDNs immediately open a CDN-website TCP connection upon receiving an HTTP/3 HEADERS frame, before the full request body arrives. This allows attackers to hold backend connections open indefinitely by slowly trickling DATA frames. We recommend three complementary strategies:
\noindent (1) \textit{Store-then-forward.} CDNs should buffer the complete HTTP/3 request (both HEADERS and DATA frames) before establishing a CDN-website connection. This is the approach already adopted by Cloudflare, which makes it immune to the HCA attack, as confirmed by our experiments (Table~\ref{table: connection time}).
\noindent (2) \textit{Limit per-connection backend fan-out.} CDNs should restrict the number of CDN-website connections that a single HTTP/3 client connection can trigger. Fastly already implements this strategy by limiting backend connections to the first 10 streams per HTTP/3 connection, which significantly reduces the HCA amplification effect.
\noindent (3) \textit{Enforce CDN-website connection timeouts.} CDNs should set strict timeout values for CDN-website connections that are independent of the client-CDN connection lifetime. For example, if no meaningful data is forwarded to the host website within 30 seconds, the CDN-website connection should be terminated. This prevents attackers from holding connections open for arbitrarily long periods (e.g., 300+ seconds in our experiments).

\subsection{Anonymity and Cost}
One may argue that launching these attacks in the real world is unlikely due to associated costs and the risk of exposing the attacker's identity. However, CDN vendors, presumably for competitive reasons, provide much convenience for their prospective customers (and thus for attackers). Table~\ref{table: anonymity and cost} shows the registration information required to begin using the free or free-trial services of the CDN vendors in our study. Three out of six CDN vendors (Baidu, Cloudflare, Fastly) require only a valid email address. CloudFront requires a valid credit card (could be a gift card or stolen). Tencent requires a valid phone number (could be anonymous or disposable). Alibaba requires users to verify their identity through a valid debit card, which takes an attacker more effort to keep anonymous. Moreover, all six vendors offer free or free-trial services, and none of them (except Cloudflare's domain blacklist) verify website ownership, meaning an attacker can register a victim's website on the CDN without the victim's knowledge. Even if CDNs widely adopted website ownership verification, an attacker could still attack any host website that enables HTTP/3 and is hosted on a vulnerable CDN, since such host websites can be targeted directly without registering the victim's domain. Combined with the minimal bandwidth required on the attacker side (less than 500Kbps for HBA with dynamic tables), the total attack cost is negligible.

\begin{table}[t]
\centering
\small
\caption{CDN registration requirements, cost, and website verification.}
\begin{threeparttable}
\begin{tabular}{@{}l@{\extracolsep{\fill}}ccc@{}}
\toprule[1.5pt]
                & \textbf{Requirements} & \textbf{Price} & \textbf{Website Verification} \\
\midrule
\midrule
\textbf{Alibaba}        & C1, C2, C4                        & Free trial     & No Verification              \\
\textbf{Baidu}          & C1                        & Free trial   & No Verification              \\
\textbf{Cloudflare}     & C1                        & Free service   & Domain Blacklist             \\
\textbf{CloudFront}     & C1, C3                        & Free trial     & No Verification              \\
\textbf{Fastly}         & C1                        & Free service   & No Verification              \\
\textbf{Tencent}          & C2                      & Free trial   & No Verification             \\
\bottomrule[1.5pt]
\end{tabular}
\begin{tablenotes}
\item[$\dagger$] C1 means an email address is required to register an account.
\item[$\ddagger$] C2 means a phone number is required to register an account.
\item[$\rVert$] C3 means a credit card is required to register an account.
\item[$\ast$] C4 means a debit card is required to register an account.
\end{tablenotes}
\end{threeparttable}
\label{table: anonymity and cost}
\end{table}

\section{Related Work}\label{sec:related_work}

We describe three categories of related work: CDN security, amplification attacks, and HTTP/3 and QUIC security.

\subsection{CDN Security}

Content Delivery Networks (CDNs) are essential to the modern Internet, handling nearly a fifth of all web traffic. Their critical role has made their security a major research focus. 
While CDNs are often lauded for their DDoS protection~\cite{imthiyas2020ddos, jalalpour2018security}, their complex infrastructure and diverse implementations present numerous vulnerabilities that attackers can exploit. This has made studying attacks on CDNs and their host websites a key topic in network security.
Previous studies have revealed a wide range of security issues in CDN ecosystems. Previous researchers have explored TLS key management issues on CDN platforms, such as private key sharing and inefficient revocation~\cite {https_meets_cdn, Private_Key_Sharing}. Furthermore, attackers can exploit inconsistencies in the interpretation of HTTP header fields between CDNs and websites to manipulate caching mechanisms, leading to cache poisoning~\cite{HoT}, cache-poisoned Denial-of-Service (CPDoS)~\cite{CPDoS}, and web cache deception (WCD) attacks~\cite{WCD}. In parallel, researchers have shown that the high reputation and relative invisibility of CDNs can be abused to circumvent Internet censorship, through techniques such as domain fronting~\cite{domain_fronting}, domain borrowing~\cite{domain_borrowing}, domain takeover~\cite{lin2024detecting}, and domain shadowing~\cite{domain_shadowing}, which remain in constant competition with evolving censorship mechanisms.

\subsection{Amplification Attack}
\noindent Amplification attacks constitute a well-established area of research within the realm of cybersecurity. Triukose et al.~\cite{2009_decopule_attack} proposed an attack that exhausts the host website's bandwidth by quickly disconnecting the client-CDN connection. Chen et al.~\cite{cdn_loop} demonstrated that inconsistencies in HTTP request handling policies across CDN vendors could be abused to create request loops among CDN nodes, repeatedly processing malicious traffic and degrading availability. Li et al.~\cite{RangeAmp} demonstrated that amplification attacks exploiting the HTTP Range Request mechanism can achieve amplification factors as high as 43,000 times. 
Guo et al.~\cite{cdn_judo} found that the compressed headers can be amplified when HTTP/2 downgrades to HTTP/1.1, which causes an amplification attack.
Guo et al.~\cite{CDN-Convex} further uncovered architectural flaws that allow adversaries to orchestrate pulse-style attack waves against websites through CDN infrastructures. Lin et al.~\cite{lin2024unveiling} revealed that attackers can abuse format transformation in CDN to drain both website and CDN bandwidth. Lin et al.~\cite{cdn_cannon} further uncovered that CDNs' back-to-origin strategies can be exploited to achieve amplification factors over 100,000.

While previous CDN amplification attacks can achieve very high amplification factors, they typically depend on specific configurations or resources on the victim's side, which significantly limits their real-world attack surface. For example, RangeAmp~\cite{RangeAmp} requires the victim website to enable HTTP Range Request support and host large files (e.g., video or software packages) on the host website; without these prerequisites, the attack cannot be launched. Similarly, CDN Cannon~\cite{cdn_cannon} achieves its high amplification factor (over 100,000$\times$) by relying on the victim enabling image optimization features on the CDN and hosting large images (e.g., 4K images) on the host website. CDN Judo~\cite{cdn_judo} exploits the HTTP/2-to-HTTP/1.1 header compression conversion, achieving amplification factors of approximately 44$\times$ (static table) and 166$\times$ (dynamic table). Our HBA attack leverages the analogous HTTP/3-to-HTTP/1.1 conversion but achieves significantly higher amplification factors of up to 66$\times$ (static table) and 350$\times$ (dynamic table), more than doubling CDN Judo's results through optimized exploitation of the QPACK dynamic table.

In contrast, our attacks impose \textit{no special configuration requirements} on the victim. The only prerequisite is that the victim website is hosted on a CDN with HTTP/3 support enabled. Critically, as shown in Table~\ref{tab: support http/3}, HTTP/3 is enabled by default on major CDN providers such as Cloudflare and CloudFront, meaning that victims may be vulnerable without any explicit action. This fundamental difference in attack prerequisites results in a significantly broader real-world attack surface: our measurement identifies 42,330 potentially vulnerable subdomains across the Tranco Top 1M, all of which satisfy the attack prerequisites without any victim-side configuration changes.

\subsection{HTTP/3 and QUIC Security}
The security of the QUIC transport protocol and HTTP/3 has attracted growing research attention. Chatzoglou et al.~\cite{quic_revisiting} presented the first comprehensive review of QUIC security and conducted fuzz testing against six production-grade QUIC servers, identifying several zero-day vulnerabilities that can lead to server resource exhaustion. In a follow-up study, Chatzoglou et al.~\cite{h3_hands_on} systematically migrated known HTTP/2 attacks to HTTP/3, testing them against six HTTP/3-enabled servers and discovering critical vulnerabilities (e.g., CVE-2022-30592). Kosek et al.~\cite{quicloris} proposed QUICLORIS, a slow-rate DoS attack targeting the QUIC protocol itself by exploiting slow connection establishment. In a subsequent work, Kosek et al.~\cite{h3_service_vuln} further analyzed security and service vulnerabilities inherent to HTTP/3 and their implications for network middleboxes.

At the transport layer, Nawrocki et al.~\cite{quicsand} quantified real-world QUIC-based reflective amplification attacks, demonstrating that QUIC servers are prone to resource exhaustion through spoofed Initial packets. Jabbari et al.~\cite{qfam} proposed QFAM, a mitigation scheme for QUIC handshake flooding attacks, showing that QUIC servers suffer a CPU amplification factor of up to 4.6$\times$ compared to TCP.

While these studies focus on vulnerabilities in QUIC implementations or the HTTP/3 protocol itself, our work addresses a fundamentally different attack surface: the \textit{HTTP/3-to-HTTP/1.1 protocol conversion} at the CDN layer. None of the above works examine how CDNs' heterogeneous protocol deployment can be exploited to amplify attacks against host websites.
\section{Conclusion}\label{sec: conclusion}

In this paper, we present the first study of DoS attacks that exploit the HTTP/3-to-HTTP/1.1 protocol conversion at the CDN layer. We design two novel amplification attacks: the HTTP/3 Bandwidth Amplification (HBA) attack, which achieves amplification factors of up to 66$\times$ with the QPACK static table and 350$\times$ with the dynamic table, and the HTTP/3 Connection Amplification (HCA) attack, which exploits HTTP/3 multiplexing combined with slow DATA frame transmission to exhaust all available connection resources of the host website. We evaluated both attacks against six major CDN providers: all six are vulnerable to the HBA attack, and five are vulnerable to the HCA attack. A large-scale measurement across the Tranco Top 1M domain list identified 151,685 subdomains hosted by vulnerable CDNs, of which 42,330 have HTTP/3 enabled and are potentially vulnerable to our attacks. The root cause lies in the heterogeneous protocol deployment where CDNs support HTTP/3 for client-CDN connections but only HTTP/1.1 for CDN-website connections. Based on this insight, we propose two design principles: bounding the amplification ratio at the conversion layer (against HBA) and validating complete requests before opening backend connections (against HCA), deployed by both Baidu and Tencent following our disclosure.
\section*{Acknowledgement}

We thank all anonymous reviewers for their valuable comments. 
This work is supported by the Science and Technology Innovation Key R\&D Program of Chongqing (No. CSTB2024TIAD-STX0022) and by the Open Topics from the Lion Rock Labs of Cyberspace Security under project LRL24006. 

\bibliographystyle{IEEEtranS}
\bibliography{references}

\end{document}